\documentclass[%
 reprint,
 amsmath,amssymb,
 aps,
floatfix,
]{revtex4-2}

\def\aap{A\&A}
\def\apj{ApJ}
\def\aapr{A\&A Rev.}
\def\apjl{ApJ}
\def\mnras{MNRAS}
\def\araa{ARA\&A}

\def\na{New Astronomy}

\def\nat{Nature}
\def\aaps{A\&A Supp.}

\def\apjs{ApJS}
\def\prd{Phys. Rev. D}

\usepackage{xspace}
\usepackage{subcaption}
\usepackage{multirow}
\usepackage{graphicx}
\usepackage{dcolumn}
\usepackage{bm}
\usepackage{hyperref}
\hypersetup{
	colorlinks=true,        
	linkcolor=blue,         
	citecolor=blue,        
	urlcolor=blue,           
}
\usepackage{dsfont}
\usepackage{comment}
\usepackage{color}
\usepackage[normalem]{ulem}
\usepackage{soul}
\usepackage{float}

\newcommand\msun{$\rm \,M_{\odot}$}

\newcommand\z{$z$}

\newcommand\mzams{$M_{\rm ZAMS}$}
\newcommand\mbh{$M_{\rm BH}$}
\newcommand{\COMPAS}{{\tt COMPAS}\xspace}
\newcommand{\MOBSE}{{\tt MOBSE}\xspace}
\newcommand{\MESA}{{\tt MESA}\xspace}

\begin{document}

\preprint{APS/123-QED}

\title{Explanation of the Mass Distribution of Binary Black Hole Mergers}

\author{Lei Li$^{1}$}

\author{Guoliang L\"{u}$^{1,2}$}%
\email{guolianglv@sina.com}
\author{Chunhua Zhu$^{1}$}
\email{chunhuazhu@sina.cn}
\author{Sufen Guo$^{1,3}$}
\author{Hongwei Ge$^{3,4,5}$}
\email{gehw@ynao.ac.cn}
\author{ Weimin Gu$^{6}$}
\author{ Zhuowen Li$^{1}$}
\author{Xiaolong He$^{2}$}

\affiliation{%
 $^{1}$School of Physical Science and Technology, Xinjiang University, Urumqi, 830046, China\\
 $^{2}$Xinjiang Astronomical Observatory, Chinese Academy of Sciences, 150 Science 1-Street, Urumqi, Xinjiang 830011, China\\
 $^{3}$Yunnan Observatories, Chinese Academy of Sciences, Kunming, 650216, China\\
 $^{4}$Key Laboratory for Structure and Evolution of Celestial Objects, Chinese Academy of Sciences, Kunming 650216, China\\
 $^{5}$International Centre of Supernovae, Yunnan Key Laboratory, Kunming 650216, China\\
 $^{6}$Department of Astronomy, Xiamen University, Xiamen, Fujian 361005, China
}%

\date{\today}

\begin{abstract}
Gravitational wave detectors are observing an increasing number of binary black hole (BBH) mergers, revealing a bimodal mass distribution of BBHs, which hints at diverse formation histories for these systems. Using the rapid binary population synthesis code \MOBSE, we simulate a series of population synthesis models that include chemically homogeneous evolution (CHE). By considering metallicity-specific star formation and selection effects, we compare the intrinsic merger rates and detection rates of each model with observations. We find that the observed peaks in the mass distribution of merging BBHs at the low-mass end (10\msun) and the high-mass end (35\msun) are contributed by the common envelope channel or stable mass transfer channel (depending on the stability criteria for mass transfer) and the CHE channel, respectively, in our model. The merger rates and detection rates predicted by our model exhibit significant sensitivity to the choice of physical parameters. Different models predict merger rates ranging from 15.4 to $96.7\,\rm{Gpc^{-3}yr^{-1}}$ at redshift $z$ = 0.2, and detection rates ranging from 22.2 to 148.3$\mathrm{yr^{-1}}$ under the assumption of a detectable redshift range of $z \le$ 1.0.

\end{abstract}

\maketitle

\section{Introduction}

Since gravitational waves (GWs) were detected for the first time in 2015 \citep{2016PhRvL.116f1102A}, the LIGO-Virgo-KAGRA (LVK) collaboration has now observed several hundred GW events, with the majority originating from mergers of binary black hole (BBH) systems \citep{2019PhRvX...9c1040A,2021PhRvX..11b1053A,2023PhRvX..13d1039A,2025arXiv250818082T}. These gradually increasing numbers of GW events have unveiled the substructure of the mass distribution of merging binary compact objects, thereby facilitating a deeper understanding of the formation and evolution of binary stars.

Extensive analyses of the BBH mass distribution in the GWTC-3 catalog using diverse modeling approaches, including parametric models \citep{Abbott2021ApJ...913L...7A,2023PhRvX..13a1048A}, autoregressive processes \citep{Callister2024PhRvX..14b1005C}, splines \citep{Edelman2023ApJ...946...16E}, piecewise “binned” models \citep{Fishbach2020ApJ...891L..31F,Veske2021ApJ...922..258V,Toubiana2023MNRAS.524.5844T}, 
binned Gaussian processes \citep{Ray2023ApJ...957...37R}, Gaussian mixture models \citep{Tiwari2021ApJ...913L..19T,Tiwari2022ApJ...928..155T}, and additive kernel density \citep{Sadiq2022PhRvD.105l3014S} all robustly identify a low-mass peak near $\sim$ 10 \msun\ and a high-mass peak near $\sim$ 35\msun. However, the feature of $\sim$ 17\msun\ has only been confirmed in studies (e.g, \citet{Tiwari2021ApJ...913L..19T,Tiwari2022ApJ...928..155T,Toubiana2023MNRAS.524.5844T}), many of them conclude that the intermediate feature is not statistically significant.

In general, these results suggest BBH systems may originate from different formation channels. \citet{Schneider2021A&A...645A...5S,Schneider2023ApJ...950L...9S} and \citet{Disberg2023} find that binary-stripped stars can explain the lower black hole (BH) mass spectrum. Population synthesis studies by \citet{vanSon2022ApJ...940..184} revealed that BBHs formed via the stable mass transfer channel can reproduce the observed peak around 10\msun\ in the detected mass distribution. The peak at 35\msun\ may originate from pulsational pair-instability supernovae \citep[PPISN][]{Belczynski2020A&A...636A.104B,2023MNRAS.523.4539K}, double-core evolution \citep{2023A&A...671A..62Q}, stable mass transfer \citep{Lucas2024MNRAS.535.2041D}, the failed supernova (FSN) mechanism \citep{Disberg2023}, early BHs produced by Population III stars \citep{2017Inayoshi,Costa2023,Santoliquido2023}, and chemically homogeneous evolution (CHE) of rapidly rotating stars \citep{Marchant2016,Mandel2016MNRAS.458.2634M,Mink2016MNRAS.460.3545D}, or comes
from a combination of CHE and stable mass transfer \citep{Briel2022MNRAS.514.1315B}. Meanwhile, it remains unclear how much each specific formation channel contributes to BBH mergers. For instance, \citet{2023MNRAS.526.4130H} argue that high-mass signatures are unlikely to originate from PPISN.
Recently, based on the GWTC-4 gravitational-wave event catalog, \citet{Banagiri2025arXiv250915646B} suggest that binary black hole merger events consist of at least three distinct subpopulations, each characterized by different primary black hole mass ranges, mass ratio distributions, and spin magnitude distributions. \citet{Roy2025arXiv250701086R} compare the observational results with various mainstream formation channels and find that none could fully explain all the observed features at 35\msun.


Besides, many studies have attempted to constrain the population properties of BBH mergers through population synthesis simulations \citep{Giacobbo2018,Belczynski2020ApJ...905L..15B,Shao2021,2023MNRAS.524..426I} or, conversely, use GW events to constrain uncertain aspects of binary evolution \citep{2021MNRAS.505.3873B}. \citet{Farmer2020ApJ...902L..36F} pointed out that if the high-mass feature indeed originates from PPISN, then the astrophysical S-factor at 300 keV of the $\rm ^{12}C(\alpha,\gamma)^{16}O$ reaction could be constrained to $S_{300} >$ 175 keV at 68\% confidence. Unfortunately, producing a satisfactory match to the observed mass distribution is challenging \citep{Stevenson2022MNRAS.517.4034S}.

Stars undergoing CHE burn nearly all their hydrogen into helium during the main-sequence (MS) phase. As a result, they retain significantly more mass at the time of collapse, leading to the formation of heavier BHs \citep{paper1,Sharpe2024ApJ...966....9S}. Consequently, they are naturally expected to be a prominent source of the high-mass end. Meanwhile, when stars undergo CHE due to rapid rotation, their radii do not expand during the MS stage due to the absence of a hydrogen envelope \citep{2006A&A...460..199Y,Mink2009A&A...497..243D,2018ApJ...858..115A,Ghodla2023MNRAS.518..860G}. This case allows them to evolve into BHs without interacting with a companion, even when the orbital separation is close, and merge within the Hubble time \citep{Marchant2016}. \citet[][]{Riley2021} incorporate the CHE channel into rapid population synthesis code \COMPAS \citep{2017NatCo...814906S,2022ApJS..258...34R} and find that it may contribute up to $\sim$ 70\% of BBH merger detections from isolated binary evolution. The detection rate may be overestimated, but it also indicates that the BBH merger rate through the CHE channel in cosmology may account for a non-negligible portion of the total merger rate \citep{2020MNRAS.499.5941D,Mandel2022LRR....25....1M}. Additionally, \citet{Dorozsmai2024MNRAS.527.9782D} investigated the evolution of hierarchical triple systems where the inner binary undergoes CHE. They found that such systems produce configurations that cannot be predicted through isolated binary evolution channels.

The primary focus of this work is to investigate the contributions of different formation channels to merging BBHs by incorporating CHE evolution from the \COMPAS code into the \MOBSE framework. The structure of this paper is as follows. In Section \ref{sec:methods}, we describe our model setup and the methods employed. Our results are presented in Section \ref{sec:results}, followed by a summary in Section \ref{subsec:Conclusion}.

\section{Methods}\label{sec:methods}

\subsection{Chemically homogeneous evolution}

To determine whether a star undergoes CHE, \COMPAS compares the rotational angular velocity during the zero age main sequence (ZAMS) with the critical rotation threshold for CHE, which was calculated beforehand using \MESA \citep{Paxton2011,Paxton2013,Paxton2015,Paxton2018,Paxton2019,Riley2021}. Following \citet{Riley2021}, values of the rotational threshold for CHE are  implemented:
$$
\omega_{M, Z}=\frac{\omega_{M, Z_{0.004}}}{0.09 \ln \left(\frac{Z}{0.004}\right)+1},
$$

where

$$
\omega_{\mathrm{M}, \mathrm{Z}_{0.004}}= \begin{cases}\sum_{i=0}^5 a_i \frac{\mathrm{M}^i}{\mathrm{M}^{0.4}} & \mathrm{rad} \mathrm{~s}^{-1}, \mathrm{M} \leq 100 \mathrm{M}_{\odot}, \\ \sum_{i=0}^5 a_i \frac{100^i}{\mathrm{M}^{0.4}} & \mathrm{rad} \mathrm{~s}^{-1}, \mathrm{M}>100 \mathrm{M}_{\odot},\end{cases}
$$

and

$$
\begin{aligned}
	& a_0=5.7914 \times 10^{-4} \\
	& a_1=-1.9196 \times 10^{-6} \\
	& a_2=-4.0602 \times 10^{-7} \\
	& a_3=1.0150 \mathrm{e} \times 10^{-8} \\
	& a_4=-9.1792 \times 10^{-11} \\
	& a_5=2.9051 \times 10^{-13}.
\end{aligned}
$$

Subsequently, the radii of CHE stars remain constant during their MS phase, and they directly evolve into helium stars upon the termination of MS. These CHE binary systems predominantly originate from overcontact binaries, which are born in close separation and are already in overcontact at the ZAMS and then subsequently undergo mass equalization, resulting in binary systems with a mass ratio q=1. At such small orbital separations, both component stars attain rapid rotational velocities due to tidal synchronization \citep{Hut1981A&A....99..126H}.

Theoretical studies focusing on stable massive overcontact binaries suggest these systems should rapidly achieve mass ratio equilibrium and subsequently evolve on nuclear timescales \citep{Kuiper1941ApJ....93..133K,Marchant2016,Menon2021MNRAS.507.5013M}. However, observations reveal a striking discrepancy: most observed massive overcontact binaries are found in unequal-mass configurations \citep{Popper1978ApJ...220L..11P,Leung1978ApJ...222..924L,1999A&AS..134..327D,Hilditch2005MNRAS.357..304H,Penny2008ApJ...681..554P,Lorenzo2014A&A...572A.110L,Mahy2020A&A...634A.119M,Janssens2021A&A...646A..33J}. \citet{Masih2022A&A...666A..18A} investigated six massive overcontact binary stars and found that these unequal-mass systems may not equalize mass as expected. Recently, \citet{Masih2025CoSka..55c.390A} conducted a review of 13 massive contact binaries and analyzed their mass ratio distribution. The results show that the mass ratios are fairly well distributed between 0.3 and 1. Although there is indeed an overdensity at q = 1, it is not as extreme as predicted by theoretical studies. 

Meanwhile, several studies find that the orbital periods of massive contact systems evolve steadily on nuclear timescales, independent of their mass ratios \citep{2014NewA...31...32Y,Li2022ApJ...924...30L,Masih2022A&A...666A..18A,Vrancken2024A&A...691A.150V}. These findings indicates that the mass ratio evolution of massive contact binaries proceeds more slowly than theoretically predicted, and may never reach a mass ratio of 1 before merging. \citet{Fabry2023A&A...672A.175F,Fabry2025A&A...695A.109F} propose that differing total masses in contact binary systems produce surface temperature differences, driving heat transfer from the hotter to the cooler stellar component. This thermal exchange promotes temperature equilibration. The resulting radius adjustments subsequently modify mass transfer rates between components, thereby influencing the mass ratio evolution. \citet{Vandersnickt2025A&A...695A.223V} investigate an alternative physical process and find that by limiting core rejuvenation through convective core overshooting, the predicted mass ratio distribution shows significant deviation from unity.

Based on the above considerations, in our toy model, we have modified the existing treatment of mass equalization in overcontact binaries to reduce the discrepancy between models and observations. We attempted to constrain the mass ratio distribution of CHE binaries in population synthesis using the limited observational data currently available for overcontact binaries. Specifically, we assumed that the mass transferred during the overcontact phase does not exceed 5\% of the secondary mass.

\subsection{Binary Population Synthesis Simulations}\label{subsec:bps}

To evolve a population of binary systems and obtain the mass distribution of merging BBHs, we use the rapid population synthesis code \MOBSE\ \citep{Giacobbo2018MNRAS.474.2959G} to evolve $\rm 10^{7}$ models. The initial parameters of the binaries are set based on observational results for massive binaries, with the initial mass distribution of the more massive primary derived from the initial mass function of \citet{Kroupa2001MNRAS.322..231K}, $p(M_{\rm 1}) \varpropto M_1^{-2.3}$. The mass ratio distribution following \citet{Sana2012}. We set the minimum mass for the primary and secondary to 5\msun, as we are only interested in BBH systems. For the initial separation, we adopt the distribution depicted by \citet{Sana2012}. We assume that the initial parameter distributions are independent of each other, and although they may be correlated \citep{Moe2017ApJS..230...15M}, this does not introduce significant uncertainty \citep{Klencki2018A&A...619A..77K}. In addition, we adopt the non-conservative mass transfer scheme proposed by \citet{Shao2016ApJ...833..108S}. It is assumed that the mass accretion rate of the rotating star is the mass transfer rate multiplied by a factor (1 - $\Omega/\Omega_{\rm crit}$), where $\Omega$ is the angular velocity of the accretor and $\Omega_{\rm crit}$ is its critical value.

To evaluate the stability of the mass transfer, the $\zeta$-prescription \citep[e.g.][]{Ge2010ApJ...717..724G,Pavlovskii2017MNRAS.465.2092P,Han2020RAA....20..161H} involves contrasting the radial responds of the donor ($\zeta_{\ast}$) with how the Roche lobe radius response to the mass transfer ($\zeta_{L}$).
In binaries that have not experienced CHE, if both components simultaneously fill their Roche lobes, or if the adiabatic expansion to mass loss of donor at the onset of mass transfer is faster than the expansion of its Roche lobe, $\zeta_{\ast}$ $<$ $\zeta_{L}$, leading to dynamical instability, the system will enter a common envelope (CE) phase \citep{Ivanova2013}. To facilitate rapid evolutionary calculations, population synthesis codes typically employ a critical mass ratio $q_{\mathrm{c}}$ \citep{Hurley2002MNRAS.329..897H}, which is calculated based on detailed evolutionary models and depends on the mass and radius of the donor star. When Roche lobe overflow commences, if the mass ratio of the binary exceeds $q_{\mathrm{c}}$, the mass transfer is unstable.

However, comparing $\zeta_{\ast}$ and $\zeta_{L}$ only at the onset of MT may not accurately predict of MT stability, as both values fluctuate throughout the process \citep{Ge2010ApJ...717..724G,Ge2015ApJ...812...40G}. \citet{Ge2020ApJ...899..132G,Ge2020ApJS..249....9G,Ge2023ApJ...945....7G,Ge2024ApJ...975..254G} and \citet{Zhang2024ApJS..274...11Z} used a detailed one-dimensional adiabatic mass loss model to explain the evolution of stability during MT events. They provided the mass ratio threshold $q_{\rm c,Ge}$ for dynamical timescale mass transfer across different evolutionary phases of the realistic donor. They found that in the new criterion $q_{\rm c,Ge}$, stars MT are more stable because the extended envelope is very diffuse, thereby limiting the increase of the mass transfer rate to a critical value. As a result, a reduced number of convective envelope donors experience dynamical instability, resulting in an increased number of systems that undergo stable MT \citep[e.g,][]{Willcox2023ApJ...958..138W}. We will discuss the impact of using $q_{\rm c,Ge}$ on the mass distribution in Section \ref{subsec:Stability}.

The CE evolution is parameterized via the classical energy prescription $\alpha_{\rm CE}$-$\lambda$\citep{1984ApJ...277..355W}, where the outcome depends on the initial binding energy of the envelope and the initial orbital energy of the binary system. Here, $\alpha_{\rm CE}$ represents the fraction of orbital energy consumed during CE evolution that contributes to envelope ejection, while $\lambda$ denotes the binding energy factor (typically $\lambda$=0.1). If orbital energy were the sole energy source for envelope ejection, $\alpha_{\rm CE}$ would be $\le$ 1. However, studies suggest additional energy sources, such as recombination energy \citep{Ivanova2015MNRAS.447.2181I,Kruckow2016A&A...596A..58K} or accretion-driven jets \citep{Soker2019MNRAS.484.4972S} may contribute. We therefore adopt $\alpha_{\rm CE}$=1.0 as our fiducial value, and also explore $\alpha_{\rm CE}$ = 0.5 and 2.0 to examine how different $\alpha_{\rm CE}$ values influence population synthesis results.

Following \citet{Belczynski2010ApJ...714.1217B}, the metallicity-dependent mass loss rate for Wolf-Rayet (WR) stars can be expressed as:
\begin{equation}
	\dot{M}_{\rm WR} = f_{\rm WR}\times 10^{-13} \left(\frac{Z}{Z_\odot}\right)^\alpha \left(\frac{L}{L_\odot}\right)^{1.5}  M_\odot\,\mathrm{yr}^{-1}
\end{equation}
where $\alpha = 0.86$ quantifies the metallicity sensitivity.
When the star is radiation-pressure-dominated ($\Gamma_e \sim$ 1), the metallicity dependence of mass loss nearly vanishes. Therefore, when $\Gamma_e \geq 2/3$, $\alpha = 2.45 - 2.4\Gamma_e$ \citep{Giacobbo2018MNRAS.474.2959G}.
Previous studies have shown that mass loss during the Wolf-Rayet phase of massive stars significantly affects BH mass \citep{Belczynski2010ApJ...714.1217B,Stevenson2022MNRAS.517.4034S}, while recent work suggests that the mass loss rates during this phase may have been overestimated in the past \citep{Sander2020MNRAS.499..873S}. Thus, the WR mass loss multiplier $f_{\rm WR}$ in our default model is set to 1.0, and we will conduct comparisons using values of 0.5 and 2.0.

The natal kick imparted during a supernova explosion can alter the orbital separation of a binary system \citep{Katz1975}. Specifically, the magnitude and direction of the natal kick received by the secondary component can significantly influence whether the binary system will merge within the age of the Universe. In wide binary systems, if the natal kick of the secondary is of appropriate velocity and direction, it may result in the secondary being drawn sufficiently close to the BH formed from the primary, thereby enabling a merger BBH within a Hubble time.
Nevertheless, the SN kick that a BH receives at birth is not precisely known. The current practice for core-collapse supernovae involves inferring the natal kick amplitudes from a Maxwellian velocity distribution, guided by measurements of radio pulsar proper motions, and assuming a one-dimensional root mean square velocity dispersion of $\rm \sigma_{rms}^{1D}$ = 265 km s$^{-1}$ \citep{Hobbs2005}, and scaled down according to the fraction of mass, $\rm f_b$, that falls back onto the newly formed compact object \citep{Fryer2012ApJ...749...91F}. Here, we adopt the natal kick prescription from \citet{Giacobbo2020ApJ...891..141G}, where the kick velocity scales proportionally with the ejected mass and inversely with the remnant mass and is able to account for both large velocities in young
isolated pulsars and small kicks in ultra-stripped SNe, electron-capture SNe, and failed SNe. To further investigate the role of natal kicks in shaping the BH mass distribution, we choose $\rm \sigma_{rms}^{1D}$ = 45 km s$^{-1}$ and 750 km s$^{-1}$ to represent cases of weak and strong natal kicks, respectively. 

\subsection{Cosmic Integration}\label{subsec:cosmic}
In essence, the BBH merger rates necessitate the consideration of two primary factors. Firstly, the distribution of delay times introduces complexity. While binaries typically evolve into BBHs within $\sim$ 10 Myr, the merger delay times can span many Gyr \citep[e.g.,][]{Peters1964,Mapelli2017MNRAS.472.2422M,Eldridge2019MNRAS.482..870E}, making it difficult to determine accurately the redshift \z\ at where the BBH systems formed. As the star formation rate density (SFRD) varies with redshift \citep{Madau2014}, this uncertainty contributes to the imprecision in predicting BBH merger rates \citep{vanSon2022ApJ...931...17V}. 
Secondly, the BBH formation rate is metallicity-dependent, with a higher yield in low-metallicity environments \citep{Giacobbo2018MNRAS.474.2959G,Giacobbo2018,Broekgaarden2022MNRAS.516.5737B}. Besides, metallicity also affects the mass loss rate, significantly impacting the remnant masses and the merger rate \citep{Langer2012,Boesky2024}. 

Based on this, our study adopts the metallicity-specific SFRD ($Z_{\rm i}, z$) from \citet{Neijssel2019}. They decomposed the calculation of the SFRD ($Z_{\rm i}, z$) into two independent factors: the SFR and the metallicity density function (${\rm d}P$/${\rm d}Z$), such that:
\begin{equation}
	\frac{\mathrm{d}^3 M_{\mathrm{SFR}}}{\mathrm{~d} t_s \mathrm{~d} V_c \mathrm{~d} Z}(z)=\frac{\mathrm{d}^2 M_{\mathrm{SFR}}}{\mathrm{~d} t_s \mathrm{~d} V_c}(z) \times \frac{\mathrm{d} P}{\mathrm{~d} Z}(z) .
\end{equation}
In this framework, the SFRD follows the parametric form proposed by \citet{Madau2014}:
\begin{equation}
	\frac{\mathrm{d}^2 M_{\mathrm{SFR}}}{\mathrm{~d} t_s \mathrm{~d} V_c}=a \frac{(1+z)^b}{1+[(1+z) / c]^d} \mathrm{M}_{\odot} \mathrm{yr}^{-1} \mathrm{Mpc}^{-3} ,
\end{equation}
with a = 0.01, b = 2.6, c = 3.2, and d = 6.2. Here, we adopt the model from \citet{Madau2017ApJ...840...39M} to characterize the SFRD. Meanwhile, for the metallicity density function (${\rm d}P$/${\rm d}Z$), we adopted the phenomenological “preferred” model from \citet{Neijssel2019}.

By applying the SFRD ($Z_{\rm i}, z$) with sampling weights adjusted according to the assumed cosmic metallicity distribution, the BBH merger rate density $\mathcal{R}_{\mathrm{m}}$, is obtained by integrating the BBH formation rate density over all metallicities:
\begin{equation}
	\begin{aligned}
		& \mathcal{R}_{\mathrm{m}}\left(t_{\mathrm{m}}, M_{\mathrm{l}}, M_2\right) \equiv \frac{d^4 N_{\text {merger }}}{d t_{\mathrm{m}} d V_{\mathrm{c}} d M_1 d M_2}\left(t_{\mathrm{m}}, M_{\mathrm{l}}, M_2\right) \\
		& =\int d Z_{\mathrm{i}} \int_0^{t_{\mathrm{m}}} d t_{\text {delay }} \operatorname{SFRD}\left(Z_{\mathrm{i}}, z\left(t_{\text {form }}=t_{\mathrm{m}}-t_{\text {delay }}\right)\right) \\
		& \times \frac{d^4 N_{\text {form }}}{d M_{\mathrm{SFR}} d t_{\text {delay }} d M_1 d M_2}\left(Z_{\mathrm{i}}, t_{\text {delay }}, M_{\mathrm{l}}, M_2\right),
	\end{aligned}\label{equ:Rm}
\end{equation}
where $V_{\mathrm{c}}$ is the comoving volume, and formation time $t_{\rm f}$ is related to the merger
time $t_{\rm m}$ and delay time $t_{\rm delay}$ by $t_{\rm form}=t_{\rm m}-t_{\rm delay}$. The latter part of Equation \ref{equ:Rm} represents the formation rate of merger BBHs per unit stellar mass formed, evaluated at a given metallicity $Z_{\mathrm{i}}$.

Following \citet{Neijssel2019}, the local detection rate $\mathcal{R}_{\mathrm{det}}$ for the Cosmic Integration pipeline\footnote{Available at \url{https://COMPAS.readthedocs.io/en/latest/pages/User guide/Post-processing/notebooks/CosmicIntegration.html}.} is given by
\begin{equation}
	\begin{aligned}
		& \mathcal{R}_{\mathrm{det}}\left(t_{\mathrm{det}}, M_1, M_2\right) \equiv \frac{d^3 N_{\mathrm{det}}}{d t_{\mathrm{det}} d M_1 d M_2} \\
		& =\int d z \frac{d V_{\mathrm{c}}}{d z} \frac{d t_{\mathrm{m}}}{d t_{\mathrm{det}}} \mathcal{R}_{\mathrm{m}}\left(t_{\mathrm{m}}\right) P_{\mathrm{det}}\left(M_1, M_2, z\left(t_{\mathrm{m}}\right)\right),
	\end{aligned}
\end{equation}
where $t_{\mathrm{det}}$ is the time  in the detector frame, $z$ is the redshift, $V_{\mathrm{c}}$ is the comoving volume, and $\mathcal{R}_{\mathrm{m}}$ is the merger rate from equation \ref{equ:Rm}.
$P_{\mathrm{det}}$ is the probability of detecting a gravitational wave signal from a binary with component masses $M_{\rm 1}$ and $M_{\rm 2}$ that merges at redshift $z$. Based on \citet{Barrett2018}, the detectability of GW signals is estimated by examining whether the signal-to-noise ratio (SNR) of the source within a single detector exceeds a predefined threshold. The SNR is calculated by computing the source waveform using a combination of the LAL suite software packages IMRPHENOMPV2\citep{2014PhRvL.113o1101H,2016PhRvD..93d4006H,2016PhRvD..93d4007K} and SEOBNRv3\citep{2014PhRvD..89h4006P,2017PhRvD..95b4010B}. Furthermore, employing the O3 sensitivity of the LIGO interferometers\citep{2020LRR....23....3A}, we assume that a GW signal can be detected when the SNR exceeds a threshold of 8.

\section{Results}\label{sec:results}
Using the \MOBSE population synthesis code, we have computed a comprehensive set of stellar population models. In this section, we first examine a ``fiducial'' model to analyze its fundamental population characteristics. Subsequently, following the methodology outlined in Section~\ref{subsec:cosmic}, we systematically calculate and discuss the merger rates and detection rates for each individual model in our parameter space study.

\begin{figure*}[ht]
	\includegraphics[width=1.\textwidth]{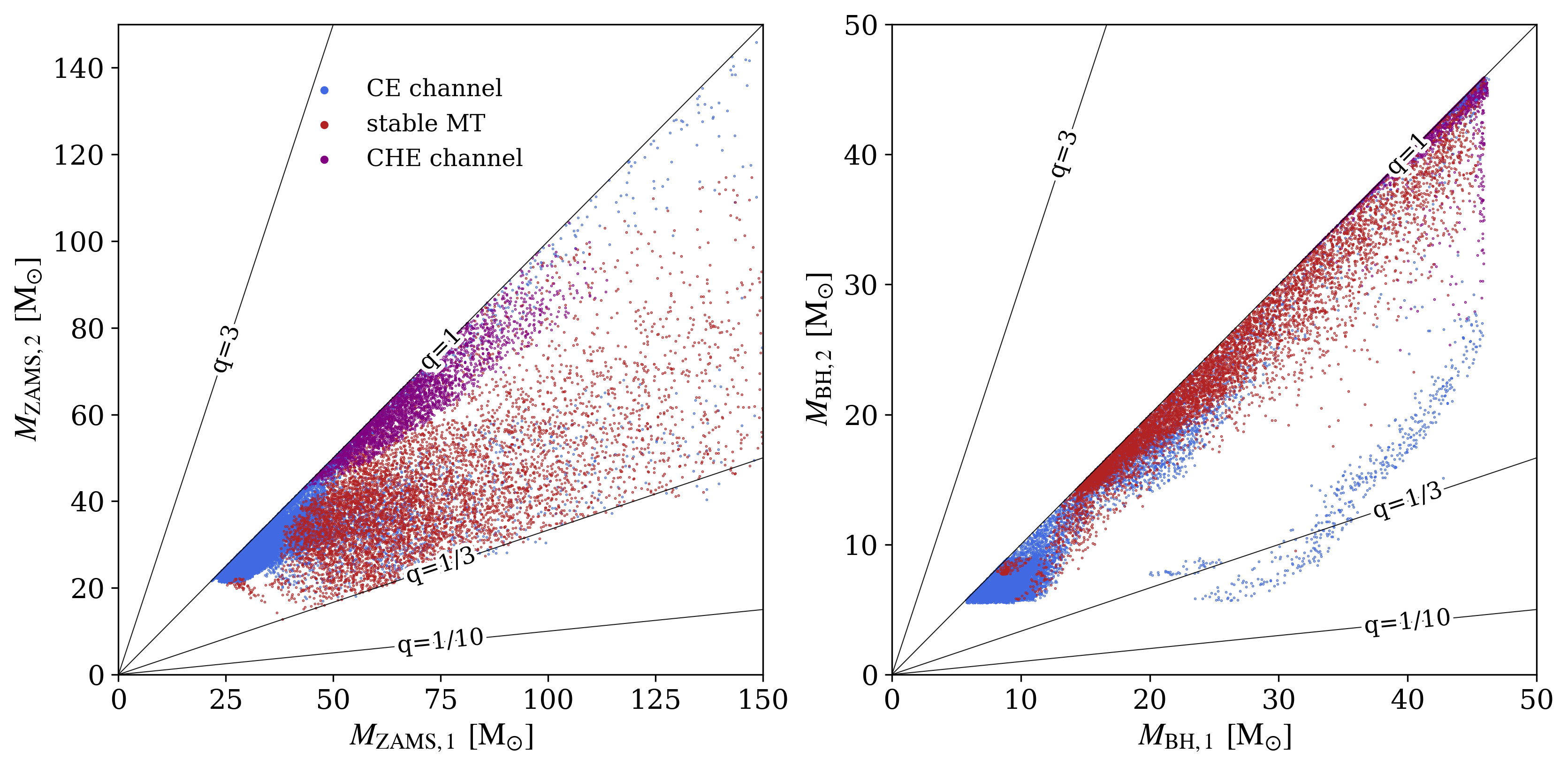}
	\caption{The distribution of different formation channels for all BBH systems in the parameter space. Different formation channels are distinguished by different colors: purple, blue, and red represent BBHs formed via the CHE channel, the CE channel, and the stable mass transfer, respectively. The black straight lines indicate the mass ratios $M_{\rm 2}$/$M_{\rm 1}$. To facilitate comparison with observational data, we designate the more massive BH as $M_{\rm BH,1}$ in this context. 
		\label{fig:channel}}
\end{figure*}
\subsection{Formation channels} \label{subsec:channel}

Based on different binary interaction processes, we classify BBH systems into three formation channels. The evolution of such binary systems must undergo at least two interaction phases, each triggered by envelope expansion of either the primary or secondary star. If both mass transfer phases remain dynamically stable, the evolutionary channel is categorized as the ``stable MT channel". If the second mass transfer is unstable, causing the binary to enter the CE phase and successfully survived, the system is categorized under the ``CE channel". Alternatively, if the initial orbital separation is sufficiently tight, tidal synchronization can spin up both stars to rotation fast enough to induce CHE. This process bypasses subsequent mass transfer interactions, placing such systems within the ``CHE channel".

{\bf Chemically homogeneous evolution channel:}
As shown in Figure \ref{fig:channel},  only stars with \mzams\ greater than 40\msun\ undergo CHE, leading to BH formation with \mbh\ generally exceeding 30\msun. 
On the other hand, the maximum BH mass is approximately 45\msun, which can be attributed to the predicted mass gap from the pair-instability supernova (PISN) and PPISN prescriptions \citep{Heger2003ApJ...591..288H,Spera2017MNRAS.470.4739S,Woosley2017ApJ...836..244W}.

{\bf Common envelope channel:}
BBHs formed from binary systems that have not experienced CHE generally have too wide separations to allow for a merger within a Hubble time, except for cases where the separation becomes sufficiently close during interactive process. In the classical CE channel, binary systems may form with relatively wide initial separations (several AU). The initially more massive star expands and fills its Roche lobe, initiating a stable MT phase with the less massive companion. In most cases, the donor loses its entire hydrogen-rich envelope during MT, while the accretor gains significant mass, potentially leading to mass ratio reversal. The donor then undergoes iron core collapse to form the first compact object. When the secondary subsequently expands and initiates MT, the system typically exceeds the $q_{\rm c}$ threshold, resulting in unstable MT and then CE evolution, which tightens the binary separation by more than two orders of magnitude \citep{Ivanova2013}. 

Figure \ref{fig:channel} also illustrates the parameter space for both the CE channel (blue dots) and the stable mass transfer channel (red dots). The BHs produced by binary systems undergoing CE evolution typically cluster around 10\msun, with their initial masses generally ranging between 20-50\msun. Additionally, we note the presence of some outlier BBH systems characterized by significantly smaller mass ratios, with some reaching q $<$ 1/3. These systems originated from binary stars with initially comparable masses, where both components underwent Roche lobe overflow during their evolution, leading to contact and subsequent CE evolution. In modeling these systems, the MOBSE code adopts an approach where the envelope of the primary is stripped while the secondary retains its envelope. This treatment ultimately produces these outlier systems post-CE evolution.

The CE channel rarely produces systems with \mbh\ greater than 30\msun, primarily because the strong stellar wind of massive progenitors is not conducive to the initiation and ejection of the common envelope, as most of the hydrogen-rich envelope is removed before interaction with the companion \citep{Mapelli2022MNRAS.511.5797M,Briel2022MNRAS.514.1315B,Belczynski2022ApJ...925...69B,vanSon2022ApJ...931...17V,vanSon2022ApJ...940..184}.

{\bf Stable mass transfer channel:} 
Furthermore, stable mass transfer can also produce tight BBH systems, especially in non-conservative scenarios \citep{Marchant2021}. Recent theoretical investigations indicate that mass transfer in massive binaries is more stable than earlier estimates \citep{Ge2020ApJ...899..132G,Ge2020ApJS..249....9G}, implying that the contribution of stable mass transfer to GW sources might be considerable. 
Consistent with the findings of \citet{Broekgaarden2021MNRAS}, BBHs formed through the stable mass transfer channel originate from binary systems with relatively compact initial separations, typically featuring initial semi-major axes smaller than 0.5 AU. However, these separations remain sufficiently wide to avoid tidal-induced CHE. These BBHs typically exhibit higher masses compared to those formed through the CE channel, with a pronounced concentration around approximately $M_{\rm BH,1}$ $\sim$ 17\msun.

\subsection{Merger delay times} \label{subsec:delay}
\begin{figure}[ht]
	\includegraphics[width=.5\textwidth]{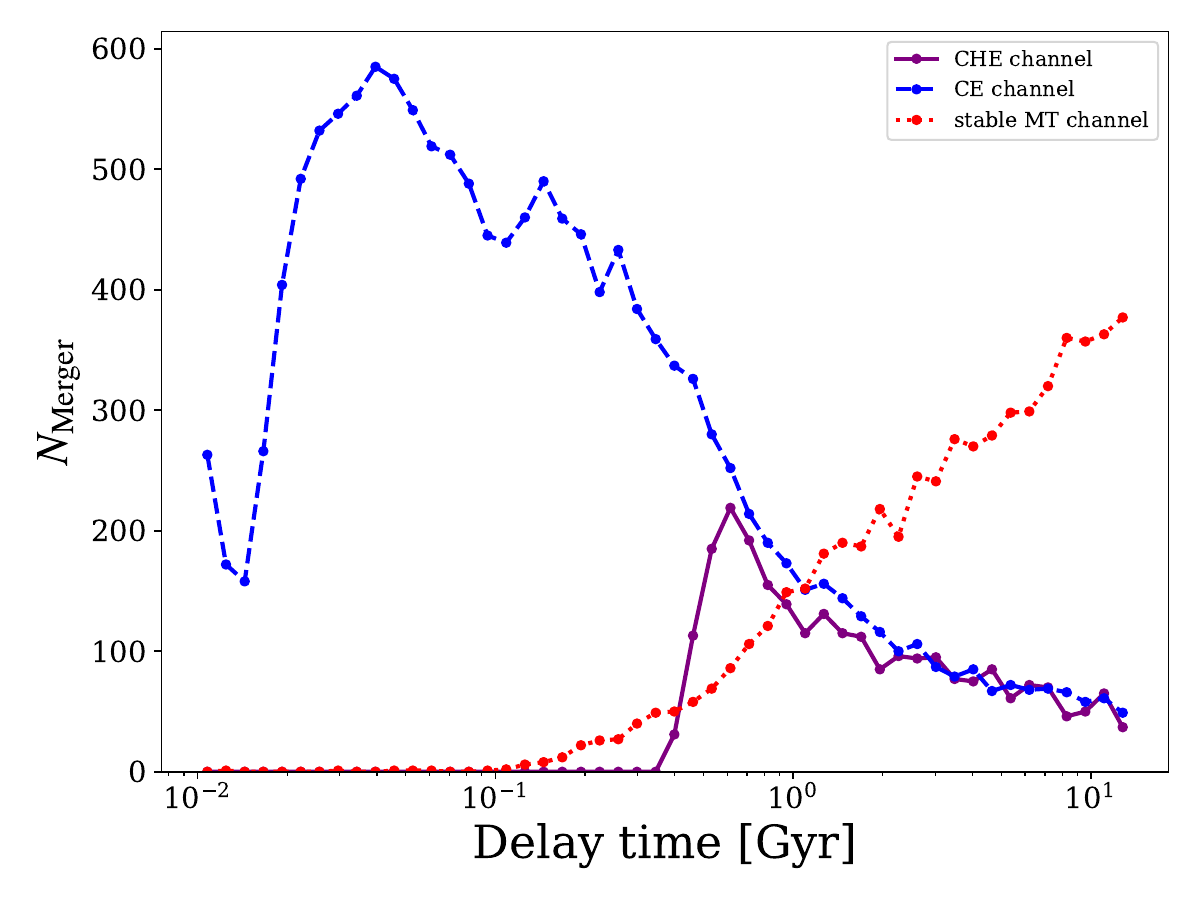}
	\caption{Distribution of delay times between formation and merger (log-scale, in Gigayears). Color-coding corresponds to different formation channels, matching the scheme in Figure \ref{fig:channel}. 
		\label{fig:delay_time}}
\end{figure}
Given the potentially substantial delay between BBH formation and coalescence, the distribution of delay times has a moderate effect on the redshift-dependent evolution of the BBH merger rate. As illustrated in Figure \ref{fig:delay_time}, the delay time distributions of BBHs vary significantly across formation channels. Notably, the CE channel has the broadest distribution, extending from as short as ~10 Myr to much longer timescales. This channel also exhibits a bias toward shorter delays, which can be attributed to the second MT phase. Since the accretor in this phase is a BH, the accretion rate is constrained by the Eddington limit, leading to efficient angular momentum removal and subsequent orbital tightening. 
Conversely, systems formed through the stable MT channel are characterized by significantly wider orbits when BBH is formed \citep{vanSon2022ApJ...931...17V}. This larger separation at formation directly translates to longer inspiral timescales during the gravitational wave emission phase, consequently producing a distribution skewed toward longer delay times.

On the other hand, the CHE channel produces a notably different delay time distribution, characterized by a sharp concentration around 600 million years. This distinctive feature arises because CHE binaries maintain nearly constant orbital separations throughout their evolution. After both stars complete their CHE phase, the binary system undergoes nearly conservative evolution with negligible angular momentum loss. Additionally, the resulting massive black holes experience only modest natal kicks, allowing the system to retain its original configuration until gravitational wave emission begins.
\subsection{Intrinsic merger rate} \label{subsec:merger_rate}
\begin{figure}[ht]
	\includegraphics[width=.5\textwidth]{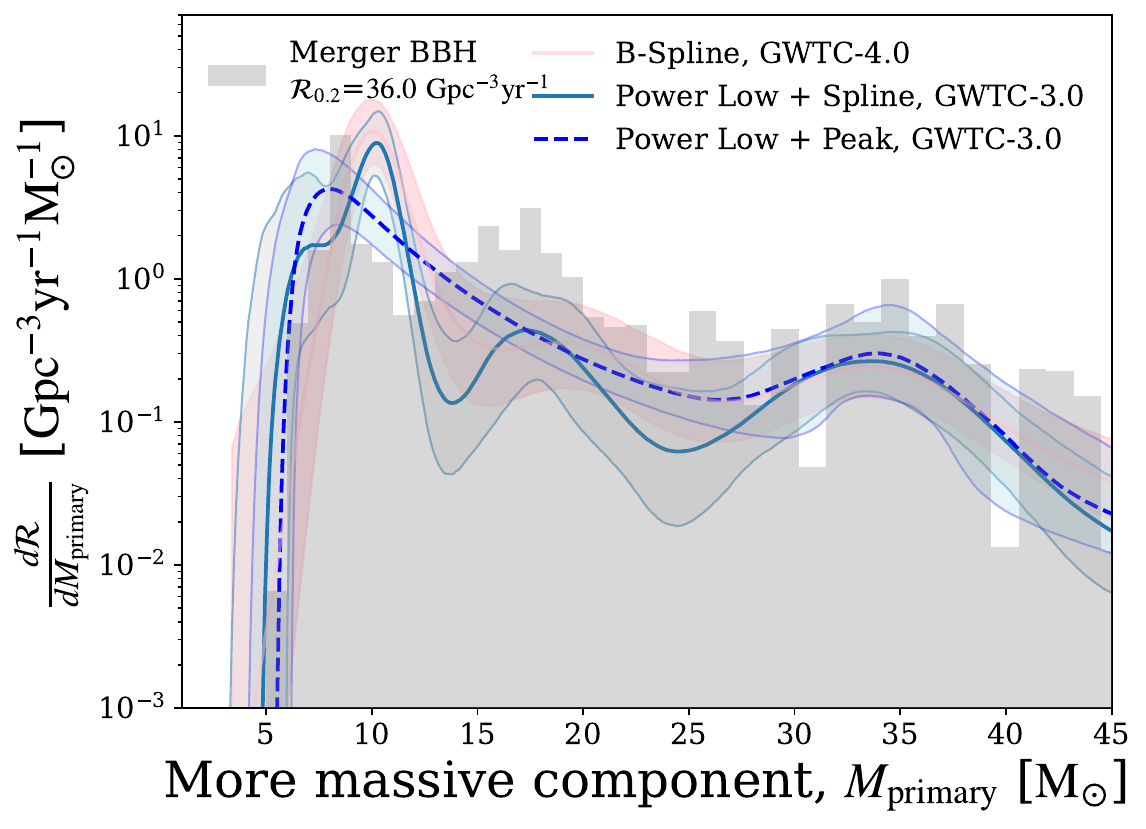}
	\caption{
		The distribution of merger rate density (per unit solar mass) at redshift $z$ = 0.2 as a function of primary BH mass in BBH mergers of fiducial model. The curves and line segments represent the posterior population distributions of the power-law + Spline (PS) model and power-law + Peak (PP) model from GWTC-3 \citep{2023PhRvX..13a1048A} and B-Spline model from GWTC-4 \citep{2025arXiv250818083T}, along with their 90\% credible intervals, indicated by the shaded regions. 
	}
	\label{fig:mc}
\end{figure}
In accordance with the methodology detailed in Section \ref{subsec:cosmic}, the intrinsic merger rate distribution of the stellar population synthesis models is computed, assuming a given SFRD ($Z_{\rm i}, z$). In Figure \ref{fig:mc}, we present fiducial model that basically reproduce the intrinsic merger rate for the primary mass predicts by \citet{2023PhRvX..13a1048A}. 
Our fiducial model essentially reproduces the peak features of GW events at 10\msun, 17\msun\ and 35\msun.
Meanwhile, our simulations produce a BBH merger rate density at $z$=0.2 of $\mathcal{R}_{0.2}$=36.0 $\rm Gpc^{-3}yr^{-1}$, which consistent with the population parameters inferred via hierarchical Bayesian analysis of the observed merger events, which is constrain the local merger rate to 17.9 $\le$ $\mathcal{R}_{0.2}$ $\le$ 44 $\rm Gpc^{-3}yr^{-1}$\cite{2023PhRvX..13a1048A}.
\subsection{Effects of input physics} \label{subsec:Uncertainties}

\begin{table*}
	\centering
	\caption{An overview of the parameters discussed in Section \ref{subsec:Uncertainties} \label{tab:uncertain}}
	\begin{tabular}{ccll}
		\hline
		Grids & Parameters & Values & Changed physics \\
		\hline
		A & $\gamma$ & [-2, -1$^{\star}$] & Angular momentum loss \\
		B & $\alpha_{\rm CE}$ & [0.5, 1.0$^{\star}$, 2.0] & CE ejection efficiency \\
		C & $f_{\rm WR}$ & [0.5, 1.0$^{\star}$, 2.0] & Mass loss multiplier \\
		D & $\rm \sigma_{rms}^{1D}$ & [45, 265$^{\star}$, 750 km $\rm s^{-1}$]  & Natal kick \\
		E & $q_{\rm c}$ & [$q_{\rm c}$$^{\star}$, $q_{\rm c,Ge}$] & Mass transfer stability  \\
		\hline
	\end{tabular}
	\smallskip
	\footnotesize
	\\The values marked with $^{\star}$ are those chosen for our basic model, which is also described in Section \ref{subsec:Uncertainties}. The parameter $\gamma$ represents different treatments of orbital angular momentum changes when mass is lost from the system during mass transfer. In the MOBSE code, the values -2 and -1 correspond to material loss and carries with it the specific angular momentum from the secondary and primary, respectively. The fiducial model implements \citet{Giacobbo2020ApJ...891..141G} natal kick prescription, $v_{\rm kick}\propto\sigma_{\rm rms}^{1D}m_{\rm ej}m_{\rm rem}^{-1}$, the kick velocity depends not only on the $\rm \sigma_{rms}^{1D}$ but also incorporates dependencies on both the ejected mass ($m_{\rm ej}$) and remnant mass ($m_{\rm rem}$), which additional complexity results in a non-linear relationship with $\rm \sigma_{rms}^{1D}$. For this comparative analysis, we instead adopt \citet{Hobbs2005}  simplified linear scheme to isolate $\rm \sigma_{rms}^{1D}$ effects on merger rate predictions. $q_{\rm crit}$ and $q_{\rm crit,Ge}$ represent two prescriptions for the mass transfer stability criterion; the former represents the default prescription in the MOBSE code, following \citet[][]{Hurley2002MNRAS.329..897H}, and the latter is the critical mass ratio calculated by \citet{Ge2020ApJ...899..132G,Ge2024ApJ...975..254G} using adiabatic mass loss from their standard stellar profiles. 
\end{table*}

\begin{figure*}
\includegraphics[width=1.0\textwidth]{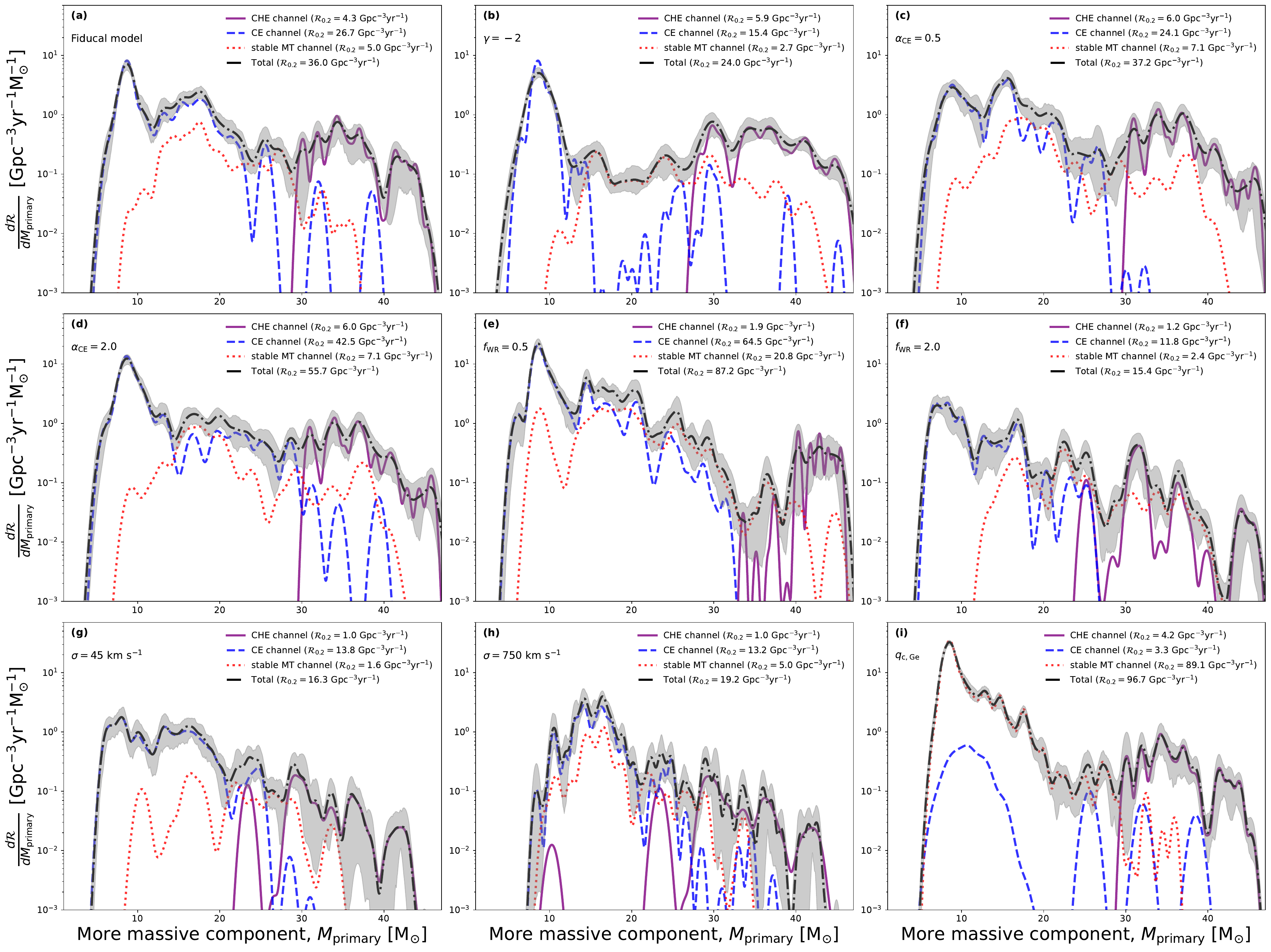}
\caption{Distribution of merger rates across different formation channels as a function of the more massive BH mass. We displays the merger rates of different formation channels at redshift $z$ = 0.2, with purple, blue, and red colors corresponding to the CHE channel, CE channel, and stable MT channel, respectively. The total merger rate at $z$ = 0.2 also included, which shown by black curve. The light-shaded area shows the 90\% sampling uncertainty as obtained from bootstrapping. Each subplot corresponds to the specific parameter configuration provided in Table \ref{tab:uncertain}, with the corresponding physical inputs indicated in the upper left corner.}  \label{fig:differ_qcrit}
\end{figure*}

In this work, we focus on the mass distribution of BBH systems formed through isolated binary evolution and achieve a distribution shape that aligns with observations. However, there are numerous uncertainties in our population synthesis calculations. We select the most significant uncertainties for discussion in this section, such as the angular momentum loss scheme $\gamma$, the CE ejection efficiency $\alpha_{\rm CE}$, the mass loss rate for WR star $f_{\rm WR}$, the natal kick and the mass transfer stability. To explore the impact of these uncertainties on the mass distribution, we design a series of models, as shown in Table \ref{tab:uncertain}.
\subsubsection{Grid A: Angular momentum loss prescriptions}\label{subsec:AM}

Panels (a) and (b) of Figure \ref{fig:differ_qcrit} respectively show the contributions from different formation channels to the merger rates for these two models. 
For fiducial model (a), the 10\msun $\,$peak is primarily governed by CE evolution, exhibiting an intrinsic merger rate density of 26.7 $\rm Gpc^{-3}yr^{-1}$, while the CHE channel primarily shapes the 35\msun feature. In comparison to the CE channel, the stable MT and CHE channels exhibit significantly lower merger rates of 5.0 and 4.3 $\rm Gpc^{-3}yr^{-1}$, respectively. 
Panel (b) presents the model of angular momentum loss originating from the primary when the binary undergoes non-conservative mass transfer. The results show that the merger rates for the stable MT channel and the CHE channel are 2.7 and 5.9 $\rm Gpc^{-3}yr^{-1}$, respectively, exhibiting minimal deviation from the baseline model. In contrast, the CE channel yields a merger rate of 15.4 $\rm Gpc^{-3}yr^{-1}$, with a pronounced decline at 17 $M_{\odot}$.

To understand why assuming mass loss from the accretor suppresses the high-mass CE channel contributions, we analyze the orbital angular momentum evolution. The change in orbital angular momentum can be conceptually understood as the combination of angular momentum lost by the donor star and angular momentum gained by the accretor. This can be expressed as:
\begin{equation}
	\dot{J}_{\mathrm{orb}} = \Delta M_1 \cdot a_1^2 \cdot \Omega_{\mathrm{orb}} + \Delta M_2 \cdot a_2^2 \cdot \Omega_{\mathrm{orb}},
\end{equation}
where $\Delta M_1$ is the mass lost by the donor, $\Delta M_2$ is the mass gained by the accretor ($|\Delta M_2| \leq |\Delta M_1|$). The orbital radii for donor and accretor are defined as
$
a_1 = \left(\frac{M_2}{M_1 + M_2}\right)a, \, a_2 = \left(\frac{M_1}{M_1 + M_2}\right)a,
$
where $a$ being the binary separation. This mechanism naturally explains the observed suppression of high-mass CE systems in model (b), as the additional angular momentum loss accelerates orbital decay, particularly for massive binaries where $a_2 > a_1$.
Given that $a_2$ invariably exceeds $a_1$, when mass is lost from the binary system, the assumption of mass removal from the accretor (as opposed to the donor star) introduces additional angular momentum loss. The stronger winds from massive stars further amplify orbital angular momentum dissipation, dramatically shrinking the orbital separation and leading to numerous binary mergers.

\subsubsection{Grids B, C: Common envelope efficiency $\alpha_{\rm CE}$ $\&$ Mass loss multiplier $f_{\rm WR}$} \label{subsubsec:CE}

There is a certain correlation between the efficiency of CE evolution ($\alpha_{\rm CE}$) and the control factor for the mass-loss rate of Wolf-Rayet stars \citep{Barrett2018}. Increasing $\alpha_{\rm CE}$ enhances the efficiency of orbital energy transfer to the CE, leading to wider post-CE orbital separations. Moreover, since mass loss through stellar wind causes the orbit to widen, increasing the mass loss rate (by raising $f_{\rm WR}$) also results in wider systems that are less likely to merge within a Hubble time. This indicates that if both parameters are increased or decreased concurrently, their effects will be amplified, but if one is increased and the other is decreased, their effects may cancel out.

In Grid B of Table \ref{tab:uncertain}, we explore the merger rate distribution for models with $\alpha_{\rm CE}$ values of 0.5 and 2.0, respectively. As shown in subplot (c) of Figure \ref{fig:differ_qcrit}, the CE channel exhibits a pronounced reduction in the merger rate peak near 10\msun. This is attributed to the lower ejection efficiency, which increases the likelihood of binary mergers during the CE phase. Conversely, for the case of $\alpha_{\rm CE}$ = 2.0 (panel d), the doubling of available ejection energy enables most binary systems to survive after the CE phase, resulting in a significant enhancement of merger rates through the CE channel. Notably, the dependence of merger rates on $\alpha_{\rm CE}$ may be non-monotonic. Some studies exploring a broader parameter space ($\alpha_{\rm CE}\in$[0.1,10.0]) report that variations in $\alpha_{\rm CE}$ can alter merger rates by up to an order of magnitude \citep{Stevenson2022MNRAS.517.4034S,Boesky2024}. Both extreme values ($\alpha_{\rm CE}$=0.1 and 10.0) yield minimal merger rates: when $\alpha_{\rm CE}$=10.0, the drastically enhanced orbital energy ejection efficiency maintains wide binary separations post-CE, preventing mergers within a Hubble time.

Panels (e) and (f) of Figure \ref{fig:differ_qcrit} demonstrate the impact of WR mass loss rates on BBH merger rates. For stronger stellar winds, merger rates across all channels decrease significantly, as enhanced mass loss widens many compact binary systems beyond the threshold for merger within a Hubble time. Conversely, when mass loss is reduced, both CE and stable MT channels show increased merger rates. However, the CHE channel maintains a low rate of only 1.9 $\mathrm{Gpc}^{-3}\mathrm{yr}^{-1}$, as a substantial fraction of stars retain massive He cores during the WR phase and subsequently undergo PISN. Furthermore, due to the exceptionally low merger rate of the CHE channel, its characteristic peak near 35\msun $\,$also disappears.

\subsubsection{Grid D: Natal Kicks} \label{subsec:kick}

Panels (g) and (h) of Figure \ref{fig:differ_qcrit} demonstrate the effect of natal kick velocity dispersion $\rm \sigma_{rms}^{1D}$ on BBH merger rates. Strong kicks can unbind binaries, significantly reducing the formation of GW sources. Panel (h) corresponds to the higher dispersion case ($\sigma_{rms}^{1D} = 750$ km/s), where most binaries are disrupted during supernovae. This results in extremely low merger rates: 8.3 $\mathrm{Gpc}^{-3}\mathrm{yr}^{-1}$ for the CHE channel, 10.4 $\mathrm{Gpc}^{-3}\mathrm{yr}^{-1}$ for the CE channel and 6.3 $\mathrm{Gpc}^{-3}\mathrm{yr}^{-1}$ for the stable MT channel.

\begin{figure*}[ht]
	\includegraphics[width=1.0\textwidth]{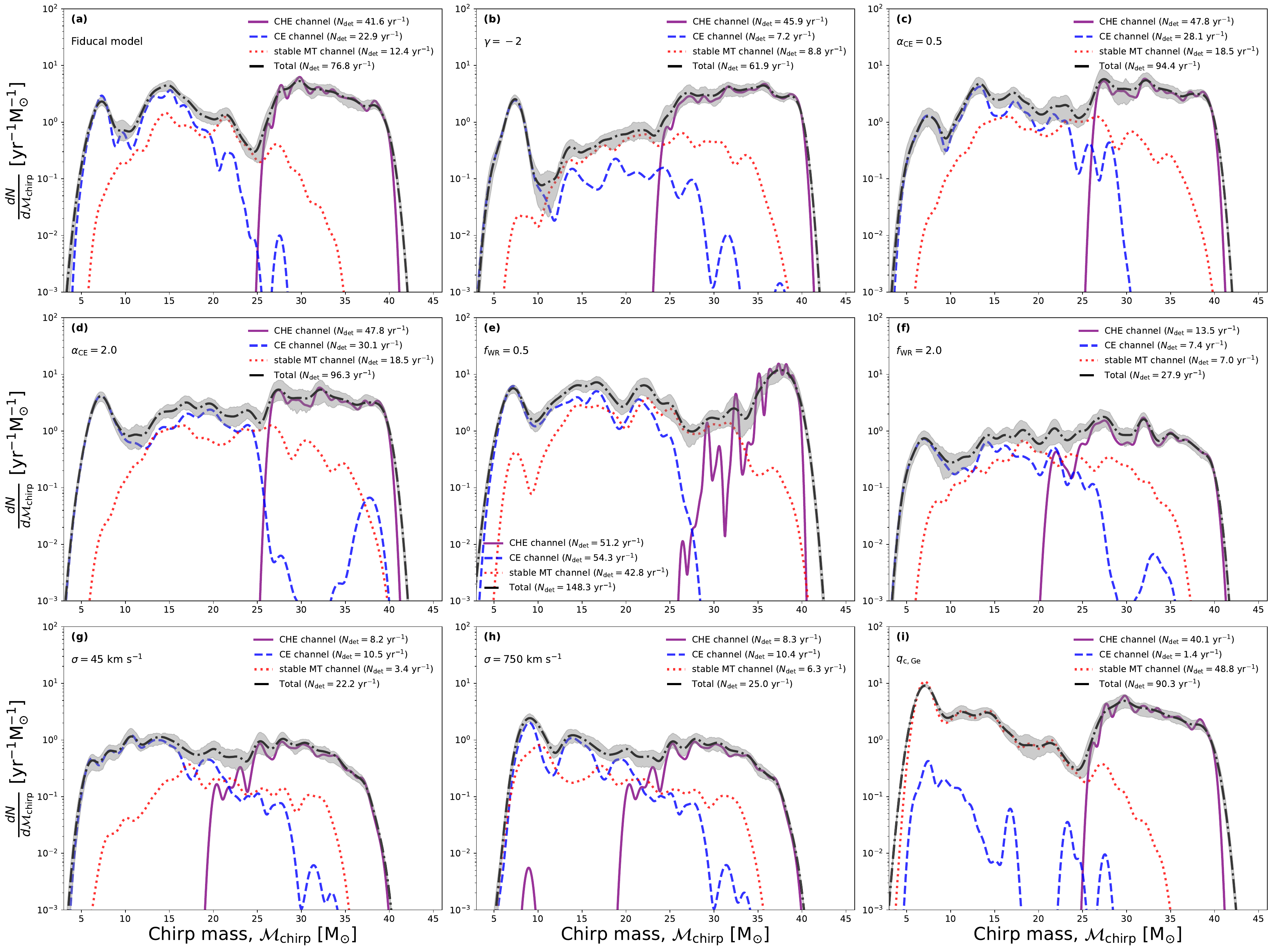}
	\caption{Distribution of detectable BBH mergers over chirp mass for differ channel, assuming LIGO O3 sensitivity, an SNR threshold of 8 and maximum detected redshift 1.0. The color-coding of lines (representing formation channels) and model numbering scheme follow the same conventions as Figure \ref{fig:differ_qcrit}. Besides, annotation labels indicate the model-specific annual detection rates (yr$^{-1}$) for each formation channel.}  \label{fig:detect}
\end{figure*}
\subsubsection{Grid E: Stability of mass transfer}\label{subsec:Stability}
If mass transfer is non-conservative, the stability criteria based on conservative mass transfer assumptions would underestimate the number of systems undergoing unstable mass transfer and stellar mergers \cite{Willcox2023ApJ...958..138W}. Therefore, we adopt the updated critical mass ratio prescription from \citet{Ge2024ApJ...975..254G}, which incorporates the effects of non-conservative mass transfer. 
Additionally, since stable thermal timescale mass transfer can lead to binary mass exchange through the outer Lagrangian point \citep{Ge2020ApJS..249....9G}, we set an upper limit of $q_{\rm c,Ge}$ ($M_{\rm donor}/M_{\rm accretor}$) = 3 for red giant branch and asymptotic giant branch stars to prevent underestimating the formation of binaries through the common envelope evolution channel.

Applying the $q_{\rm c,Ge}$ criterion, we find a fundamental shift in merger channel dominance. As shown in panel (i) of Figure \ref{fig:differ_qcrit}, this configuration results in the stable MT channel dominating the merger population – a distinctive feature not seen in other models. Crucially, the characteristic 10 M$_\odot$ peak is now primarily sustained by stable MT systems. At redshift $z=0.2$, the merger rate densities are 4.2, 3.3, 89.1 $\mathrm{Gpc}^{-3}\mathrm{yr}^{-1}$ for CHE channel, CE channel, and stable MT channel, respectively, yielding a total merger rate of 96.7 $\mathrm{Gpc}^{-3}\mathrm{yr}^{-1}$. This implies that the stable MT channel predicts an excessively high merger rate at the low-mass end. In the formation scenario, BBHs formed through the stable MT channel undergo two phases of stable mass transfer, driven by the expansion of the primary and secondary stars (referring here to the more massive and less massive stars at the time of binary formation, respectively). When Roche lobe overflow occurs, due to the larger value of updated criterion $q_{\rm c,Ge}$ compared to the classical $q_c$ value, binaries that would merge via CE under classical $q_{\rm c}$ instead evolve through stable mass transfer

\subsubsection{Chirp mass distribution of detectable BBH mergers}\label{subsec:detect}

The chirp mass 
\begin{equation}
	\mathcal{M}=\frac{\left(m_1 m_2\right)^{3 / 5}}{\left(m_1+m_2\right)^{1 / 5}},
\end{equation}
is directly related to the phase evolution of GW and is more well-constrained than the individual masses $m_{\rm1}$ and $m_{\rm 2}$ for low-mass systems that are dominated by the inspiral \citep{Peters1964,Blanchet1996CQGra..13..575B}. Figure \ref{fig:detect} displays the detection rates  over chirp mass for different formation channels under the LIGO O3 sensitivity. The predicted detection rates vary significantly across different models, with the overall detection rate ranging from 22.2 to 148.3 $\rm yr^{-1}$.

Our analysis demonstrates that the majority of detected low-mass BBH systems ($M_{\rm BH}<$ 30\msun) are likely formed through the CE evolutionary channel. Specifically, the predicted detection rates show that CE channel systems dominate the population by a factor of approximately 2 compared to those formed via stable MT across most of our considered models. This pronounced difference in formation efficiency persists throughout our parameter space studies, suggesting that CE evolution plays a crucial role in shaping the observable low-mass BBH population.
On the other hand, selection effects favoring more massive BBHs, which means that stronger GW signals are more readily detectable, lead to a higher probability of detecting BBHs formed through the CHE channel is higher. These systems dominate the predicted observable population in 50\% of our parameter configurations (Models a, b, c and d), contributing 50-74\% of the total detectable BBH mergers. Overall, in our model, the contribution of the CHE channel to observable BBHs is lower than the results from \citet{Riley2021}. Except for model b where the CHE channel contributes $74\%$ of observable BBHs, models a, c, and d exhibit contributions in the range of $50\%$--$54\%$, while the remaining models show contributions between $33\%$ and $48\%$. This discrepancy primarily arises because our model does not assume equal mass division in contact systems, leading to fewer surviving systems.

Meanwhile, our findings align with the bimodal chirp mass distribution reported by \citet{2023PhRvX..13a1048A}, which shows prominent peaks at 8\msun\  and 28 \msun, along with a weak feature at 15\msun. While our fiducial model (panel a) reproduces this triple-peak structure, the 15\msun\ feature exhibits notable parameter sensitivity. The 8\msun\ and 28\msun\ peaks consistently originate from CE and CHE channels, respectively, but the 15\msun\ peak demonstrates channel variability, alternately dominated by either CE or stable MT channels across different parameter configurations.

\section{Conclusion} \label{subsec:Conclusion}

In this work, we systematically investigate the contributions of the CE channel, stable MT channel, and CHE channel to BBH merger rates by incorporating metallicity-specific star formation. Furthermore, we employ different models to examine the effects of mass transfer stability, natal kicks, $\alpha_{\rm CE}$, and $f_{\rm WR}$ on both the merger rate and detection rate.

Our models successfully reproduce the observed peaks in the BH mass distribution at 10\msun\ and 35\msun\ (corresponding to chirp mass peaks at 8\msun\ and 28\msun). The lower-mass peak (10\msun) primarily originates from either the CE channel or the stable MT channel, depending on the critical mass ratio prescription, while the higher-mass peak (35\msun) is predominantly contributed by the CHE channel. The predicted merger rates at redshift $z=0.2$ span a range of $15.4$--$96.7\,\rm{Gpc^{-3}yr^{-1}}$.
Among the models, model (a) and model (b) exhibit the best agreement with the observed mass distribution (see Figure \ref{fig:mc}). Their respective merger rates are $36.0$ and $24.0\,\rm{Gpc^{-3}yr^{-1}}$, which lie within the observationally inferred range of $17.9 \leq \mathcal{R}_{0.2} \leq 44\,\rm{Gpc^{-3}yr^{-1}}$. Most notably, in model (i) employing the $q_{\rm c, Ge}$ prescription, while the high-mass peak remains dominated by the CHE channel, the low-mass merger rate is almost entirely governed by the stable MT channel. Besides, this model predicts an excessively high merger rate in the stable MT channel, as numerous binaries that would have been classified as undergoing unstable mass transfer under previous criteria $q_{\rm c}$ now experience stable mass transfer and avoid merge under the new prescription. On the other hand, our models predict detection rates spanning 22.2–148.3 yr$^{-1}$, with the best observationally consistent models (a) and (b) yielding rates of 76.9 and 61.9 yr$^{-1}$, respectively. Furthermore, due to selection effects, the relative contribution fraction from the CHE channel becomes significantly enhanced.


\section*{Acknowledgements}
We thank Professor Ilya Mandel for helpful discussions. We are also grateful to the anonymous reviewer for their constructive feedback. This work received the support of the National Natural Science Foundation of China under grants U2031204, 12163005, 12373038, and 12288102; the National Key R\&D Program of China under grant 2023YFA1607901; the Natural Science Foundation of Xinjiang No.2022TSYCLJ0006 and 2022D01D85; the science  research grants from the China Manned Space Project with No.CMS-CSST-2021-A10. HG acknowledges supports from Chinese Academy of Sciences (No. XDB1160201), the National Key R\&D Program of China (No. 2021YFA1600403), and NSFC No. 12173081.

\section*{Data available}
Simulations in this paper made use of the \MOBSE rapid binary population synthesis code, which is freely available at: \url{https://gitlab.com/mobse/source-code}. The code in this study are publicly available\citep{zenodo}.

\bibliography{sample631}

\end{document}